\documentclass[11pt,twoside,a4paper,twocolumn]{article}
\usepackage{graphicx}
\usepackage[utf8]{inputenc}
\usepackage{amsmath}
\usepackage{amssymb}
\graphicspath{ {.}{../images} }
\usepackage{xcolor}
\usepackage{colortbl}
\usepackage{multirow}
\usepackage{float}
\usepackage{subcaption} 
\usepackage{booktabs}  

\begin{document}

\title{A refinement of the Lorentz local field expression with impact on the Clausius-Mossotti and Lorentz-Lorenz models}
\author{Jeroen van Duivenbode$^{1}$, Anne-Jans Faber$^{2}$, Reinoud Lavrijsen$^{3}$}
\date{	
	$^1$\textit{Department of Electrical Engineering, Eindhoven University of Technology}\\
	$^2$\textit{Physica Fit Faber, Heteren, the Netherlands}\\
	$^3$\textit{Department of Applied Physics, Eindhoven University of Technology}\\
	\today \\
	}
\maketitle

\begin{abstract}
In the 19th century Mossotti and Clausius developed an expression linking the electrical permittivity of a dielectric to the product of molecular polarizability and number density. Lorenz and Lorentz later extended this framework to encompass the dielectric’s refractive index. These classical expressions have proven remarkably successful in describing how permittivity and refractive index vary with number density, under the assumption that molecular polarizability remains relatively constant.\\

While these models have stood the test of time and continue to offer valuable insights, their derivation relies on an approximation of the local electric field within a spherical cavity that simulates the molecular environment, excluding the field generated by the molecule or molecules themselves. For regimes of higher number densities, such as those encountered in densified dielectrics, employing an exact solution for the local field becomes increasingly important. This refinement extends the applicability of the Clausius-Mossotti and Lorentz-Lorenz equations and leads to more accurate estimates of molecular polarizability in general.
\end{abstract}

\section{Introduction}\label{se:1}

Many publications and textbooks, e.g. \cite{Clausius-MBdE,Lorentz-UdBzdFdLudK,FLPII-11,Griffiths-ItE, Kittel, Ashcroft}, describe the local, uniform field inside a spherical void cavity in a linear, isotropic dielectric medium using Eq. (\ref{eq:local_field_Lorentz}) in one of the following forms:
\begin{align}
	\boldsymbol{E}_\mathrm{local}&=\boldsymbol{E}+\boldsymbol{E}_1\notag\\
	&=\boldsymbol{E}+\frac{1}{3}\cdot\frac{\boldsymbol{P}}{\varepsilon_0}\notag\\
	&=\boldsymbol{E}\cdot\left(1+\frac{\varepsilon_\mathrm{r}-1}{3}\right)\notag\\
	&=\boldsymbol{E}\cdot\frac{\varepsilon_\mathrm{r}+2}{3}, 
	\label{eq:local_field_Lorentz}
\end{align}
where $\boldsymbol{E}$ is the average static field, $\boldsymbol{P}$ the average polarization, $\varepsilon_0$ the vacuum permittivity and $\varepsilon_\mathrm{r}$ the dielectric's relative permittivity. This local field, also known as the Lorentz field, is used to model the electrical environment seen by an approximately spherical individual molecule, its own field excluded (as the molecule cannot be polarized by its own field). Term $\boldsymbol{E}_1$ models the contribution from the surrounding molecules to the void left by the excluded molecule. The field of that molecule is $\boldsymbol{E}_\mathrm{net,inside}=-\boldsymbol{E}_1$, so that $\boldsymbol{E}_\mathrm{local}+\boldsymbol{E}_\mathrm{net,inside}=\boldsymbol{E}$. The local field solution has been used to derive the Clausius-Mossotti equation \cite{Clausius-MBdE, Mossotti-MDhSDdE} which links the product $N\alpha$ of molecular polarizability $\alpha$ and number density $N$ to the medium's relative permittivity $\varepsilon_\mathrm{r}$. It has also been used to derive the Lorentz-Lorenz equation \cite{Lorentz-UdBzdFdLudK, Lorenz-EoTUoLB}  that relates $N\alpha$ to refractive index $n$.\\

This paper traces all steps in establishing these equations and shows that Eq. (\ref{eq:local_field_Lorentz}) is a rough approximation. The exact solution to the local field will be shown to be
\begin{align}
	\boldsymbol{E}_\mathrm{local}&=\boldsymbol{E}+\boldsymbol{E}_1\notag\\
	&=\boldsymbol{E}+\frac{1}{2\varepsilon_\mathrm{r}+1}\cdot\frac{\boldsymbol{P}}{\varepsilon_0}\notag\\
	&=\boldsymbol{E}\cdot\left(1+\frac{\varepsilon_\mathrm{r}-1}{2\varepsilon_\mathrm{r}+1}\right)\notag\\
	&= \boldsymbol{E}\cdot\frac{3\varepsilon_\mathrm{r}}{2\varepsilon_\mathrm{r}+1}\label{eq:local_field_Maxwell},
\end{align}
which matches the field in a spherical cavity in a dielectric \cite{Onsager-EMML, Jackson-CE, Zangwill}.\\ 

\begin{figure}[h]
	\centering
	\includegraphics[width=\linewidth]{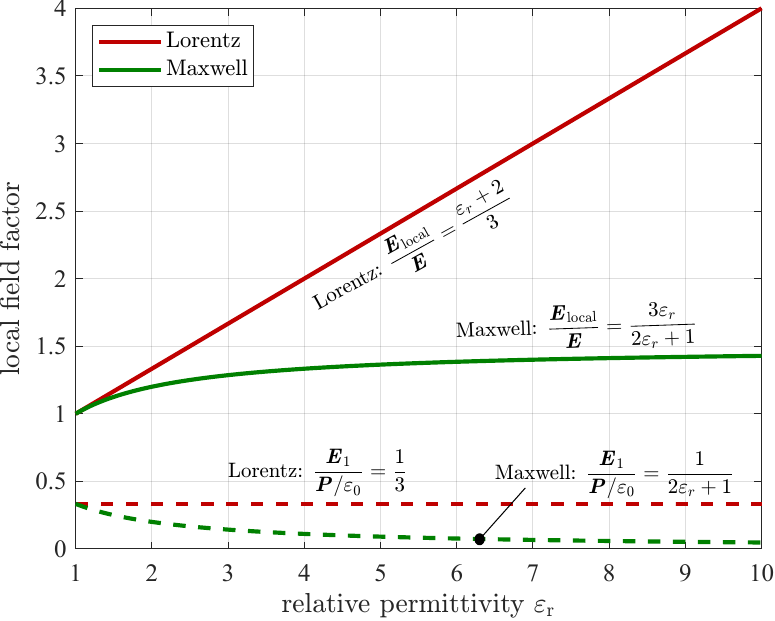}	
	\caption{Local field factors within a cavity in a dielectric, with average surrounding field $\boldsymbol{E}$ imposed. Red lines, marked "Lorentz", follow Eq. (\ref{eq:local_field_Lorentz}); green lines, marked "Maxwell", follow Eq. (\ref{eq:local_field_Maxwell}). Dashed lines indicate how polarization adds to the local field; solid lines the resulting field amplication factor. }
	\label{fig:Lorentz local field}
\end{figure}

Fig. \ref{fig:Lorentz local field} compares the field factors introduced above and shows the difference is significant. As an example, for a regular value of $\varepsilon_\mathrm{r}=4$, Eq. (\ref{eq:local_field_Lorentz}) overestimates the local field by 50\%.
\\

Section \ref{se:2} will explain the origin of this difference and establish Eq. (\ref{eq:local_field_Maxwell}) as the exact expression for the local field. Sections \ref{se:3} and \ref{se:4} recall the original Clausius-Mossotti and Lorentz-Lorenz models and relates them to the local field expression they are based on. With the exact field solution refined expressions are derived, and application examples are presented. Section \ref{se:5} shows how estimates of molecular polarizability change going from the original to the alternative expressions. An extension for the local field, from a single molecule to a spherical set of molecules, is given in Appendix \ref{app:Spherical_set}. Appendix \ref{app:Field} provides the background for the field solutions of uniform-field spheres based on a particular distribution of the sphere’s surface charge, for different sets of relative permittivity inside and outside the sphere. It then shows how the application of an external field induces such sphere polarization and provides the combined field solutions, for all sets, including generic equations which cover all of them. A verification by finite element modeling is given.

\section{The exact local field}\label{se:2}

\begin{figure*}[h]
	\centering
	\includegraphics[width=\linewidth]{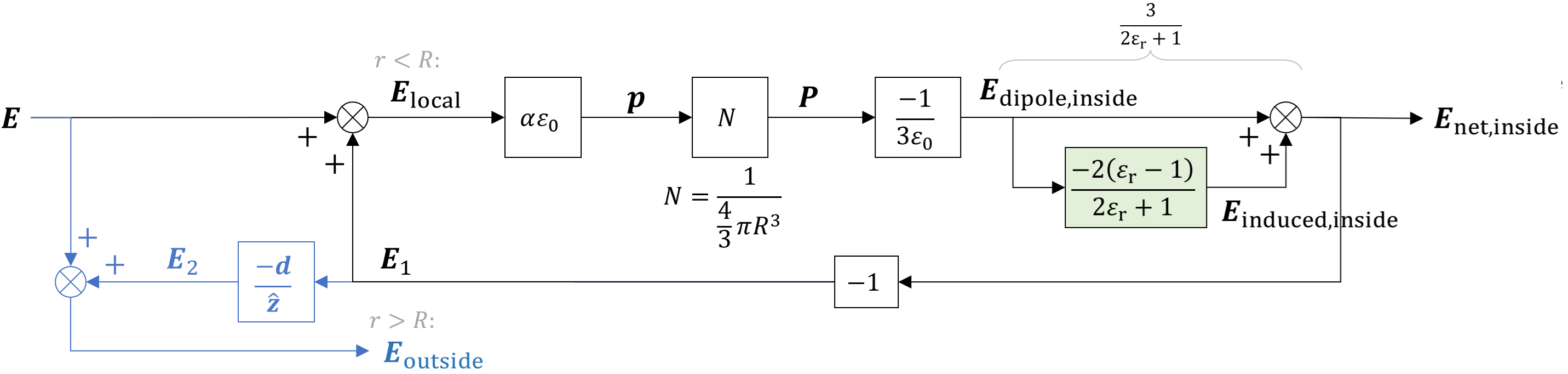}	
	\caption{Field algebra diagram with the feedback loop that links all field components inside the spherical cavity. Outside field components are added for completeness.}
	\label{fig:Field_algebra}
\end{figure*}

Figure \ref{fig:Field_algebra} shows global average field $\boldsymbol{E}=E\cdot \hat{\boldsymbol{z}}$ triggering $\boldsymbol{E}_\mathrm{local}$. This $\boldsymbol{E}_\mathrm{local}$ will induce the molecule's polarization $\boldsymbol{p}$, whose field $\boldsymbol{E}_\mathrm{net,inside}$ is subtracted from $\boldsymbol{E}$ following the logical argument that the molecule's dipole cannot be induced by the dipole itself. $\boldsymbol{p}$ is scaled by polarizability $\alpha$ and aligned with the local electric field:
\begin{equation}
	\boldsymbol{p}=\alpha\varepsilon_0\boldsymbol{E}_\mathrm{local}\label{eq:p_vs_E_local}.
\end{equation}

The molecule is modeled to occupy a spherical volume $\frac{4}{3}\pi R^3=1/N$, $N$ being the molecular number density. The molecule's induced dipole is modeled as a surface charge $P\cos{\theta}$ spread over the sphere surface\footnote{Alternatively, modeling the dipole as charge spread over a smaller sphere, or as a pure dipole at the sphere center, will not affect the end result of the local field which was the field inside the sphere, the molecule's own dipole field excluded.} leading to polarization (or polarization density):
\begin{equation}
	\boldsymbol{P}=N\alpha\varepsilon_0\boldsymbol{E}_\mathrm{local}\label{eq:P_vs_E_local}.
\end{equation}
This $\boldsymbol{P}$ must match the average polarization which, by definition of $\varepsilon_\mathrm{r}$, is:
\begin{equation}
	\boldsymbol{P}=(\varepsilon_\mathrm{r}-1)\varepsilon_0\boldsymbol{E}\label{eq:P_definition}.
\end{equation}

The $P\cos{\theta}$ surface charge generates field inside and outside the sphere. Disregarding corrective field term $\boldsymbol{E}_\mathrm{induced,inside}$ for now, one third of the generated flux goes inside and generates a uniform field $\boldsymbol{E}_\mathrm{dipole,inside}$, in the opposite direction of the polarization. Two thirds of the flux go outside to form a pure dipole field $\boldsymbol{E}_\mathrm{net,outside}=-\boldsymbol{E}_2$, modulated by the dipole field factor
\begin{equation}
	\boldsymbol{d}=\frac{2\cos{\theta}\cdot\hat{\boldsymbol{r}}+\sin{\theta}\cdot\hat{\boldsymbol{\theta}}}{(r/R)^3} \label{eq:dipole_field_factor_d},
\end{equation}
that is derived using $\boldsymbol{E}=-\nabla V$ from Eq. (\ref{eq:V_12_sigma_f_r_z}) in Appendix \ref{app:Field}. In this expression, $r$ and $\theta$ are the regular spherical coordinates with the sphere centered at the origin. $\hat{\boldsymbol{z}}$, $\hat{\boldsymbol{r}}$ and $\hat{\boldsymbol{\theta}}$ are associated unit vectors. \\

Field component $\boldsymbol{E}_\mathrm{dipole,inside}$ is negated into $\boldsymbol{E}_1$ and then added to $\boldsymbol{E}$ to form $\boldsymbol{E}_\mathrm{local}$.  Uniform field $\boldsymbol{E}_1$ is scaled by $-\frac{\boldsymbol{d}}{\hat{\boldsymbol{z}}}$ into dipole field $\boldsymbol{E}_2$ which, added to $\boldsymbol{E}$ makes $\boldsymbol{E}_\mathrm{outside}$. Difference term $\boldsymbol{E}_\mathrm{local}$ generates dipole moment $\boldsymbol{p}$ as stated above. Due to the net positive feedback, $\boldsymbol{E}_\mathrm{local}$ is stronger than $\boldsymbol{E}$.\\

Figure \ref{fig:Field_algebra}, however, also shows the corrective field term $\boldsymbol{E}_\mathrm{induced,inside}$ from the charge the dipole induces in the dielectric around it (see green-colored box). This field term weakens the effect of $\boldsymbol{E}_\mathrm{dipole,inside}$:
\begin{align}
	\boldsymbol{E}_\mathrm{net,inside}&=\boldsymbol{E}_\mathrm{dipole,inside}\cdot\left(1-\frac{2(\varepsilon_\mathrm{r}-1)}{2\varepsilon_\mathrm{r}+1}\right) \notag \\
		&=\boldsymbol{E}_\mathrm{dipole,inside}\cdot\frac{3}{2\varepsilon_\mathrm{r}+1}\label{eq:E_net_inside_from_E_dipole_inside}.
\end{align}
$\boldsymbol{E}_\mathrm{induced,inside}$ attenuates both the field inside and outside the sphere while the relative 1/3-inside, 2/3-outside flux distribution is preserved. With this attenuation, $\boldsymbol{E}_\mathrm{local}$ resolves to Eq. (\ref{eq:local_field_Maxwell}); without it, it resolves to Eq. (\ref{eq:local_field_Lorentz}).\\

Should the field the dipole induces in its environment be considered for the local field, or not? According to standard field analysis, it should - just as a point charge is attracted by the mirror charge it induces in a grounded plane; any charged object experiences not only external fields but also those it induces in its surroundings. \\

Hence, the exact local field is described by Eq. (\ref{eq:local_field_Maxwell}). This makes it equal to the field in a spherical cavity in a dielectric (i.e. without a polarized molecule in that cavity). Maxwell published that field solution in the year 1858 \cite{Maxwell-OFLF}, though for an analogous situation in a magnetic field, see Eq. (\ref{eq: Voltage cases 2 as per Maxwell}) in Appendix \ref{app:Field}. Indeed, adding to that field a pure dipole with moment $p=P\cdot \frac{4}{3}\pi R^3$  produces uniform field $\boldsymbol{E}$ outside the cavity. Adding the dipole in the form of surface charge $P\cdot\cos{\theta}$ results in field $\boldsymbol{E}$ both outside and inside the cavity.\\

In this section, $R$ has been considered to be the radius of a single molecule. Appendix \ref{app:Spherical_set} shows that the exact local field model applies equally well to a spherical set of many molecules. This makes the exact shape of a single molecule irrelevant as long as its field resembles that of a pure dipole.\\

The Clausius-Mossotti and Lorentz-Lorenz models are based on the local field concept, combined with the approximate solution of Eq. (\ref{eq:local_field_Lorentz}), i.e. neglecting corrective field term $\boldsymbol{E}_\mathrm{induced,inside}$. Section \ref{se:3} illustrates how applying the conventional approach of including this corrective term, as per Eq. (\ref{eq:local_field_Maxwell}), refines the Clausius-Mossotti model and effectively eliminates the so-called "Mossotti catastrophe". Section \ref{se:4} demonstrates how this refinement enhances the Lorentz-Lorenz model and makes it suitable even for highly densified materials.

\section{The Clausius-Mossotti model}\label{se:3}
\begin{figure}[h]
	\centering
	\includegraphics[width=\linewidth]{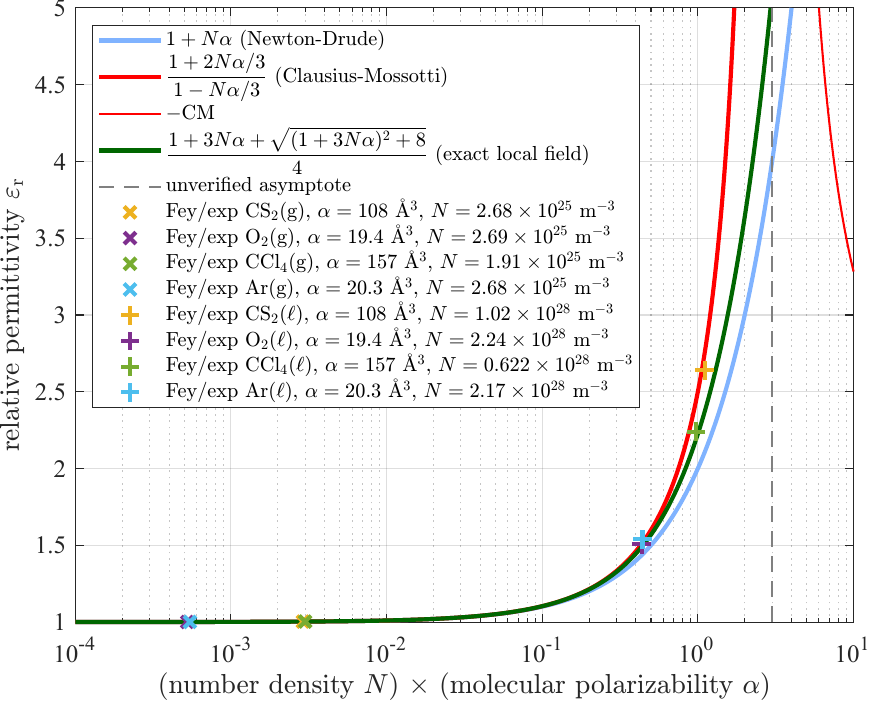}	
	\caption{Comparing experimental data with three models for relative permittivity versus the product of number density and molecular polarizability.}
	\label{fig:eps_r_vs_Nalpha}
\end{figure}

As discussed in Section \ref{se:1}, the net field from the molecule's dipole is minus the field produced by its neighbors in the situation of the void cavity. If the material is sparse, such as in a gas, the influence of polarized neighbor molecules can be neglected and $\boldsymbol{E}_\mathrm{local}\approx\boldsymbol{E}$. Then Eq. (\ref{eq:P_vs_E_local}) and (\ref{eq:P_definition}) combine to the Newton-Drude expression:
\begin{equation}
	\varepsilon_\mathrm{r}=1+N\alpha\label{eq:Newton-Drude}.
\end{equation}
Combining Eq. (\ref{eq:P_vs_E_local}) and (\ref{eq:P_definition}) with Eq. (\ref{eq:local_field_Lorentz}), the \textit{approximative} solution for $\boldsymbol{E}_\mathrm{local}$, gives the Clausius-Mossotti\cite{Clausius-MBdE} (CM) equation:
\begin{align}
	N\alpha=&3\cdot\frac{\varepsilon_\mathrm{r}-1}{\varepsilon_\mathrm{r}+2}\notag\\
	\Leftrightarrow \varepsilon_\mathrm{r}=&\frac{1+2N\alpha/3}{1-N\alpha/3}\label{eq:Clausius-Mossotti}.
\end{align}
Combining Eq. (\ref{eq:P_vs_E_local}) and (\ref{eq:P_definition}) with Eq. (\ref{eq:local_field_Maxwell}), the \textit{exact} solution for $\boldsymbol{E}_\mathrm{local}$,  gives:
\begin{align}
	N\alpha=&\frac{(\varepsilon_\mathrm{r}-1)(2\varepsilon_\mathrm{r}+1)}{3\varepsilon_\mathrm{r}}\notag\\
	\Leftrightarrow\varepsilon_\mathrm{r}=&\frac{1+3N\alpha+\sqrt{(1+3N\alpha)^2+8}}{4}\label{eq:improved_Clausius-Mossotti}.
\end{align}
Figure \ref{fig:eps_r_vs_Nalpha} compares how the three models fit to experimental data for gases and liquids\footnote{Values of $\alpha$ and $N$, reported in the legend, are derived using $M_\mathrm{CS_2}=76.13$, $M_\mathrm{O_2}=31.999$, $M_\mathrm{CCl_4}=153.82$ and $M_\mathrm{Ar}=39.948$, unit $\mathrm{g/mol}$. The values of $\alpha$ correspond fairly well with the atomic polarizability values reported in \cite{Schwerdtfeger-Nagle}, summing these values according to molecule composition and using the conversion factors given in Footnote \ref{fn:units} for the values of $\alpha$.}, taken from \cite{FLPII-11}. As expected, the Newton-Drude model is the worst fit. Based on this dataset the Clausius-Mossotti (\ref{eq:Clausius-Mossotti}) model and the one based on the exact local field solution (\ref{eq:improved_Clausius-Mossotti}) both extrapolate the gas-based data points to the liquid-based data points impressively well, but neither provides a perfect fit. This can be attributed to measurement errors in the gas-phase relative permittivity with values very close to unity, and also to the basic assumption of a molecule's spherical volume, which certainly is an approximation for the three shown species that do not have spherical symmetry. The Clausius-Mossotti model does exhibit an asymptote at $N\alpha=3$, with transition to negative values of the relative permittivity. This asymptote is known in literature as "Curie point" \cite{Onsager-EMML} or "Mossotti catastrophe" \cite{Eremin} but has not been supported by experiments. It disappears in the model based on the exact field of Eq. (\ref{eq:local_field_Maxwell}).\\

A second case study is to see how a dielectric's refractive index relates to the molecular polarizability of the constituent molecules, based on the local field models of either Eq. (\ref{eq:local_field_Lorentz}) or (\ref{eq:local_field_Maxwell}). 

\section{The Lorentz-Lorenz model}\label{se:4}
With Maxwell's identity for the refractive index $n=\sqrt{\varepsilon_\mathrm{r}\mu_r}$, where $\varepsilon_\mathrm{r}$ is the relative permittivity for electronic polarization that is applicable to the high frequencies of light, and magnetic permeability $\mu_r=1$ for most dielectrics, the Clausius-Mossotti Equation (\ref{eq:Clausius-Mossotti}), which we established to be based on Eq. (\ref{eq:local_field_Lorentz}), can be written as the Lorentz-Lorenz (LL) equation:
\begin{align}
	N\alpha=&3\cdot\frac{n^2-1}{n^2+2} \notag \\
	\Leftrightarrow n^2=&\frac{1+2N\alpha/3}{1-N\alpha/3}. \label{eq: Lorentz-Lorenz}
\end{align}
Application of the exact local field (ELF) solution as per Eq. (\ref{eq:local_field_Maxwell}) changes this to:
\begin{align}
	N\alpha=&\frac{(n^2-1)(2n^2+1)}{3n^2} \notag \\
	\Leftrightarrow n^2=&\frac{1+3N\alpha+\sqrt{(1+3N\alpha)^2+8}}{4} \label{eq: Corrected Lorentz-Lorenz}.
\end{align}
	
To compare these LL and ELF models, data is taken from Arndt-Hummel \cite{Arndt-Hummel-GRFAtDSG}, reporting measured refractive index of glass compounds with a $\mathrm{TiO_2}$ fraction of either $0.9$, $4.7$ or $7.5~\mathrm{mol\%}$, or $\mathrm{Na_2O}$ fraction of either $20$ or $33~\mathrm{mol\%}$, at different degrees of densification. This data set has significant values of refractive index, less susceptible to measurement errors than the close-to-unity values used in Section \ref{se:3}.  The larger number of data points also allows to identify trends, including the hypothesis of constant molecular polarizability and the assumption that it may be based on a linear, molar fraction based combination of the molecular polarizability of its constituents, also claimed in \cite{Arndt-Hummel-GRFAtDSG}.\\

With molecular mass $M_\mathrm{Si}=28.085$, $M_\mathrm{Ti}=47.867$, and $M_\mathrm{O}=15.999~\mathrm{g/mol}$, each compound $i$ with $\mathrm{TiO_2}$ molar fraction $k_i$, molecular mass
\begin{equation}
	M_i=(1-k_i)\cdot(M_\mathrm{Si}+2M_\mathrm{O})+k_i\cdot(M_\mathrm{Ti}+2M_\mathrm{O})\label{eq:molecular_mass_of_compound}
\end{equation}
and density $\rho_j$ will have number density:
\begin{equation}
	N_{i,j}=\frac{N_A\cdot\rho_j}{M_i}\label{eq:number density}
\end{equation}
where $N_A$ is Avogadro's number. With linear polarization the weighted compound molecular polarizability is:
\begin{align}
	\alpha_{\mathrm{c}_i}&=(1-k_i)\cdot\alpha_\mathrm{SiO_2}+k_i\cdot\alpha_\mathrm{TiO_2}\notag\\
	\Leftrightarrow \boldsymbol{\alpha}_\mathrm{c}&=\boldsymbol{K}\boldsymbol{\alpha_\mathrm{bm}}\label{eq:alpha compound},
\end{align}
where $\boldsymbol{\alpha_\mathrm{bm}}$ is a column vector with the polarizabilities of the base materials and $\boldsymbol{K}$ is a matrix with columns $(1-k_i)$ and $k_i$. Inversely, $\boldsymbol{\alpha}_\mathrm{bm}$ can be retrieved from a large number of data points with least square fitting using
\begin{equation}
	\boldsymbol{\alpha}_\mathrm{bm}=(\boldsymbol{K}^T\boldsymbol{K})^{-1}\boldsymbol{K}^T \boldsymbol{\alpha}_\mathrm{c}\label{eq:alpha_bm}.
\end{equation}
A similar apporach is used for the compounds containing $\mathrm{Na_2O}$ with $M_\mathrm{Na}=22.990$ .\\

\begin{figure}[h]
	\centering
	\includegraphics[width=\linewidth]{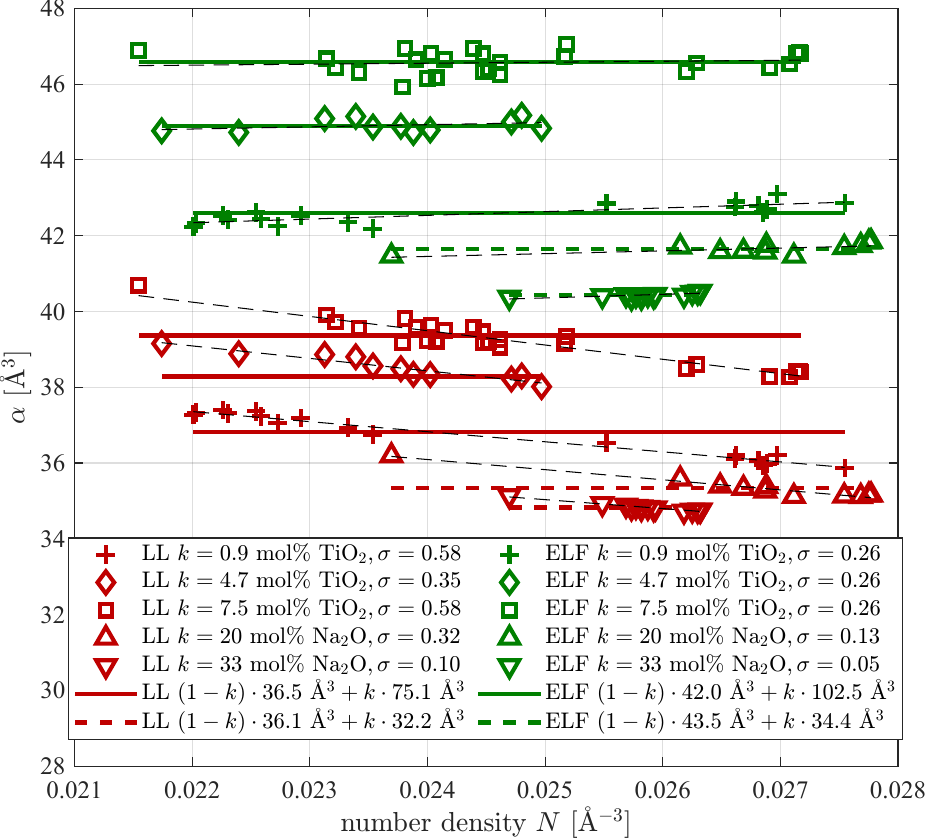}	
	\caption{Compound and base material molecular polarizability $\alpha$ recovered from refractive index values reported by Arndt-Hummel with the LL and ELF models (method discussed in Appendix \ref{app:Method}). Both models corroborate the linearity hypothesis. Only the ELF model corroborates the hypothesis of constant $\alpha$. The ELF model predicts higher values of $\alpha$.}
	\label{fig:alpha LL and ELF vs N}
\end{figure}

Fig. \ref{fig:alpha LL and ELF vs N} validates, with three different molar fractions of $\mathrm{TiO_2}$, the assumption of linearity between the compound molecular polarizability and that of its constituents. Though the data is affected by some measurement noise, it is clear that the $\alpha_\mathrm{c}$ numbers recovered using the LL model show a tilt that goes against the hypothesis of constant molecular polarizability. The numbers recovered using the ELF model corroborate this hypothesis much better which indicates that the investment in a better field model does pay off. The tilt also results in higher values of standard deviation between the data points and the expected horizontal lines of constant $\alpha$ for each compound, as reported in the legend. The difference in absolute values of $\alpha$ will be discussed in Section \ref{se:5}.\\

\begin{figure}[h]
	\centering
	\includegraphics[width=\linewidth]{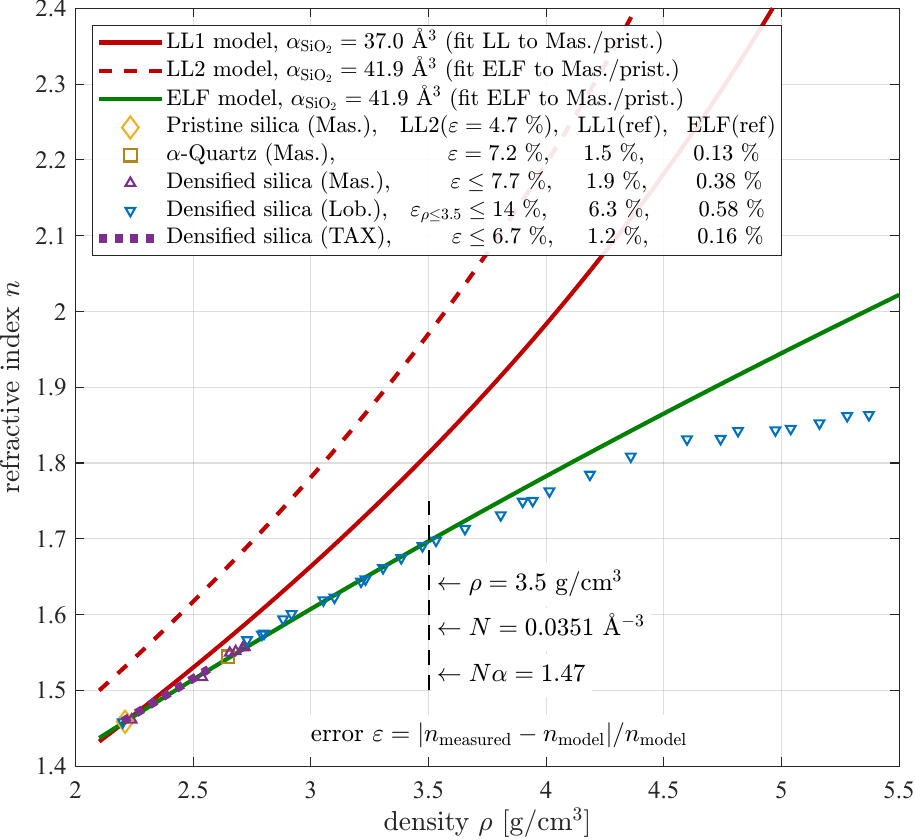}	
	\caption{Refractive index versus density with highly densified $\mathrm{SiO_2}$ measured by Lobanov\cite{Lobanov-ESMPSG} and others\cite{TanArndtXie-OPoDSG}. The ELF model supports the constant molecular polarizability hypothesis up to significant degrees of densification.}
	\label{fig:Masuno and Lobanov}
\end{figure}
Fig. \ref{fig:Masuno and Lobanov} is based on three other data sets from different sources including one with very high levels of densification. It further confirms that the ELF model corroborates the fundamental concept/hypothesis of $\alpha$ staying constant up to high densities\footnote{$\alpha$ stays constant, so density $\rho$ is proportional to $N\alpha$, which, inserted in Eq. (\ref{eq: Corrected Lorentz-Lorenz}), predicts the refractive index.}. This is validated by Lobanov's  \cite{Lobanov-ESMPSG} highly densified glasses (blue triangles) which track the green ELF model line up to $\rho=3.5 ... 4~\mathrm{g/cm^3}$.\\

Two variants of the LL model are presented. Two (different!) values of $\alpha_\mathrm{SiO_2}$ make the LL1 and ELF model curves pass through the reference data point of $n_\mathrm{pristine}=1.459$ (yellow diamond). Already for low degrees of densification with $\rho\le 3~\mathrm{g/cm^3}$, the data points quickly diverge from the LL1 model (red line, Eq. (\ref{eq: Lorentz-Lorenz})) but track the ELF model (green line, Eq. (\ref{eq: Corrected Lorentz-Lorenz})) with great accuracy. Assuming the ELF model predicts $\alpha_\mathrm{SiO_2}=41.9~\mathrm{\AA^3}$ correctly and inserting this into Eq. (\ref{eq: Lorentz-Lorenz}) gives the LL2 model (red dashed line). Error percentages (see legend) underline the better match of the test data to the ELF model.\\

\section{Absolute values of molecular polarizability}\label{se:5}
In the previous section, the curve fitting involved using the Lorentz-Lorenz and exact local field based models to recover molecular polarizability $\alpha$ of the base materials\footnote{Authors differ in units used for molecular polarizability. $1~\mathrm{a.u.}$ (atomic units) $=\mathrm{a_0}^3\approx 0.148184~\mathrm{\AA^3}$, where $\mathrm{a_0}$ is the Bohr radius.  $1~\mathrm{\AA^3}$ (Gaussian units) $=4\pi~\mathrm{\AA^3}$ (SI units). This paper uses SI units.\label{fn:units}}, presented in Table \ref{tb:Polarizability}.

\definecolor{col_LL}{rgb}{0.6, 0, 0}
\definecolor{col_ELF}{rgb}{0, 0.35, 0}
		
\begin{table*}[t]
	\centering
	\caption{Molecular polarizability in $\mathrm{\AA}^3$ (SI), derived from refractive index.}
	\label{tb:Polarizability}
	\begin{tabular}{|r|c|c|c|c|c|}
		\toprule
		& \multicolumn{3}{c|}{$\alpha_\mathrm{SiO_2}$} &  $\alpha_\mathrm{TiO_2}$ &  $\alpha_\mathrm{Na_2O}$ \\
		& Masuno & {Arndt-Hummel} & {Jeziorkowski} & {Arndt-Hummel} & {Jeziorkowski}  \\
		
		\midrule
		LL-model:  &  \textcolor{col_LL}{$37.0$} &  \textcolor{col_LL}{$36.5$} & \textcolor{col_LL}{$36.1$} & \textcolor{col_LL}{$75.1$} & \textcolor{col_LL}{$32.2$}  \\
		\midrule
		ELF-model: & \textcolor{col_ELF}{$41.9$} \textcolor{gray}{$(+13\%)$} & \textcolor{col_ELF}{$42.0$} \textcolor{gray}{$(+15\%)$} & \textcolor{col_ELF}{$43.5$} \textcolor{gray}{$(+20\%)$} & \textcolor{col_ELF}{$102.5$} \textcolor{gray}{$(+36\%)$} & \textcolor{col_ELF}{$34.4$} \textcolor{gray}{$(+7\%)$} \\
		\bottomrule
	\end{tabular}	 	
\end{table*}

Clearly the models result in different values for polarizability. The Clausius-Mossotti and Lorentz-Lorenz models are based on the concept of a molecule being polarized by the local field that reigns inside the otherwise void sphere that represents its volume. The ELF solution is shown in Fig. \ref{fig:Situation 2c}. Both the field inside the sphere and the one outside equal the undisturbed field shown in Fig. \ref{fig:Situation 2d}, minus the field in Fig. \ref{fig:Situation 1c}, with source term $\sigma_\mathrm{f_0}$ set to $P$, the polarization in the undisturbed field situation. Surface charge $P\cos{\theta}$ induces bound charge of opposite polarity in the dielectric outside the sphere, leaving a reduced net charge. However, Eq. (\ref{eq:local_field_Lorentz}), used to calculated the Clausius-Mossotti and Lorentz-Lorenz equations, subtracts the field of Fig. \ref{fig:Situation 1a} instead, thus omitting the induction of bound charge in the dielectric surface around the sphere.\\ 

As a result, even though the Clausius-Mossotti and Lorentz-Lorenz equations have been used to extrapolate relative permittivity and refraction index with reasonable accuracy (even going from gas phase to liquid phase), this extrapolation diverges from actual values when densified materials are considered, as shown in Fig. \ref{fig:Masuno and Lobanov} (difference between red line and blue data points). Also, these models underestimate the general absolute values of molecular polarizability significantly as shown in Table \ref{tb:Polarizability} and the difference between the dashed red line and the blue data points in the same graph. These values are not only applicable to densified materials but also to gases and non-densified liquids and solids.

\section{Conclusion}
Two different solutions are found in literature for the local field inside a void spherical cavity that is situated in a linear isotropic dielectric exposed to an otherwise homogeneous electric field. One is an approximation, the other one is exact. This paper establishes the exact solution and shows it is equal to Maxwell's solution for the field in a spherical void cavity. Appendix \ref{app:Field} documents two methods that resolve it. The exact solution refines two formulas that relate a dielectric's density to its relative permittivity (Clausius-Mossotti) and refractive index (Lorentz-Lorenz) respectively. The refined formulas predict these parameters more accurately, with corrections becoming substantial when densified materials are considered. Inversely, estimations of generic molecular polarizabity, based on measurements of density and relative permittivity or refractive index, interpreted with the original Clausius-Mossotti or Lorentz-Lorenz formulas, need corrections that can exceed 10\% for normal liquids and solids. Future work could use the introduced methodology and apply it to the Debye equation which also contains a permanent polarization term for polar molecules, but is based on the original Clausius-Mossotti equation. Revised values of molecular polarizability may improve models that are based on those values. Future research could apply the considerations of this paper to non-spherical molecules.

\section*{Acknowledgement}
The authors thank Bas Vermulst, Robert Jan van Wijk, Kurt Vergeer, Ton Backx, Dick Harberts, Michel de Lange, Ruben Kok, Joost Frenken, Margriet van der Heijden and Jan van Dijk for their encouragements and help on this paper.

\appendix
\section*{Appendices}
\section{Spherical set of molecules}\label{app:Spherical_set}
\begin{figure}[h]
	\centering
	\includegraphics[width=\linewidth]{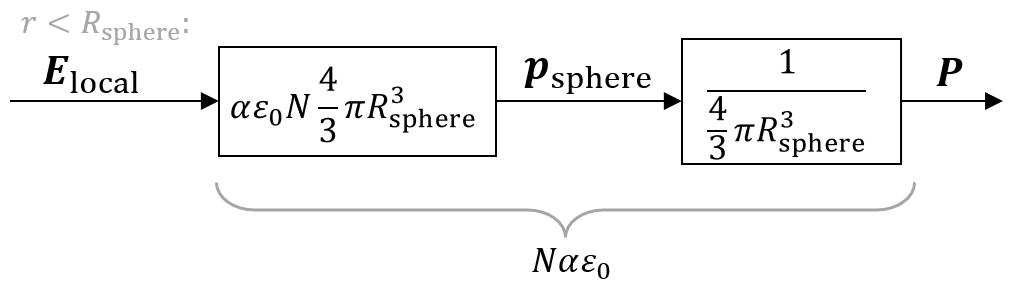}
	\caption{Alternative path from $\boldsymbol{E}_\mathrm{local}$ to $\boldsymbol{P}$ for a spherical set of molecules. The non-modified part of Fig. \ref{fig:Field_algebra} remains valid.}
	\label{fig:Spherical_set}
\end{figure}

Figure \ref{fig:Spherical_set} shows how polarization $\boldsymbol{P}$ can be linked to an alternative version of uniform local field $\boldsymbol{E}_\mathrm{local}$, where the latter is then experienced by a spherical set of molecules instead of the single molecule modeled in Fig. \ref{fig:Field_algebra}. In an intermediate step, their combined dipole moment $\boldsymbol{p}_\mathrm{sphere}$ is calculated. Division by the sphere's volume $\frac{4}{3}\pi R_\mathrm{sphere}^3$ then gives average polarization $\boldsymbol{P}$. For this spherical set of molecules, Eq. \ref{eq:P_vs_E_local} still validates the derivations of the Clausius-Mossotti and Lorentz-Lorenz models and their improved versions described in sections \ref{se:3} and \ref{se:4}. The only condition is that the combined field of the spherical set matches that of a pure dipole outside the sphere, and that of a $P\cos{\theta}$ distributed dipole surface charge inside.

\section{Method}\label{app:Method}
To produce the plot of Fig. \ref{fig:alpha LL and ELF vs N}, the data was pre-processed as follows: 1) Eq. (\ref{eq:number density}) converts the density numbers in the data to number density. 2) Eq. (\ref{eq: Lorentz-Lorenz}) and (\ref{eq: Corrected Lorentz-Lorenz}) convert the measured refraction index data into values of $N\alpha$, respectively for the original Lorentz-Lorenz model and for the exact local field based model. 3) The $N\alpha$ numbers are divided by the number density values found in the first step to give $\boldsymbol{\alpha}_\mathrm{c}$ for the different compounds. 4) Eq. \ref{eq:alpha_bm} fits $\boldsymbol{\alpha}_\mathrm{bm}$ to the values of the previous step.\\

\section{Field of uniform-field spheres}\label{app:Field}
\begin{figure*}[h!]
	\centering
	\begin{subfigure}[t]{0.49\textwidth}
		\centering
		\includegraphics[height=7.5cm]{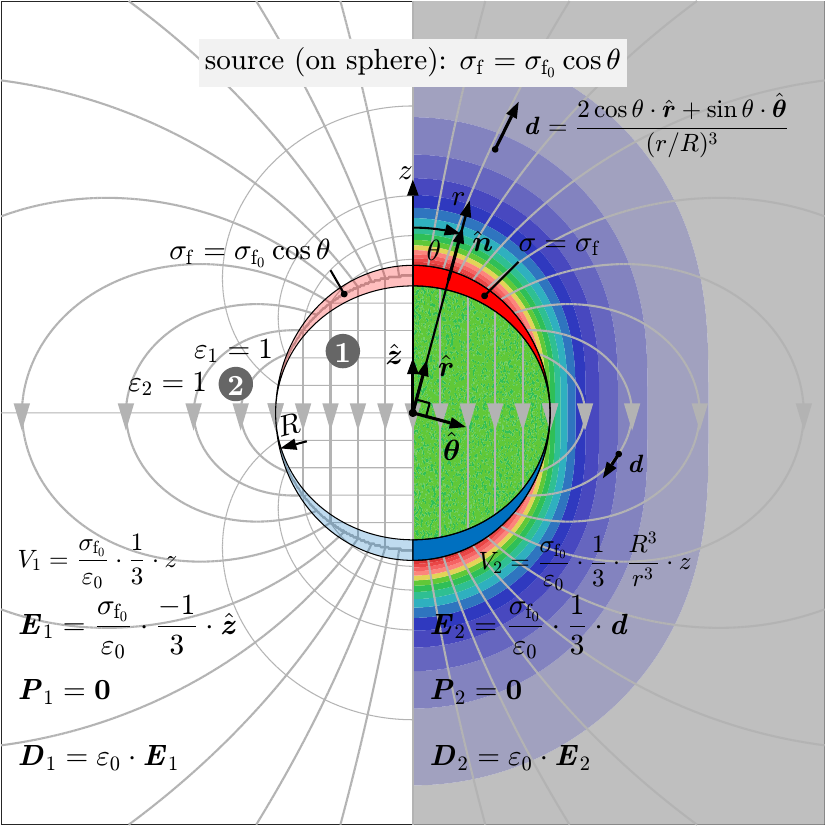}
		\caption{in vacuum}	
		\label{fig:Situation 1a}
	\end{subfigure}%
	~
	\begin{subfigure}[t]{0.49\textwidth}
		\centering
		\includegraphics[height=7.5cm]{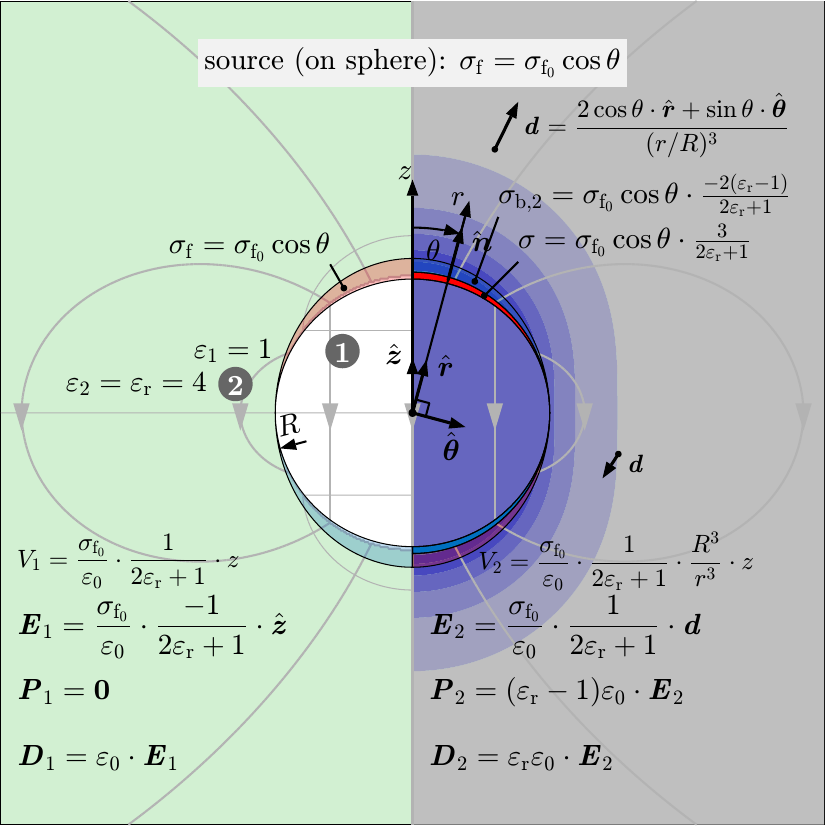}
		\caption{in dielectric}
		\label{fig:Situation 1c}	
	\end{subfigure}
	\caption{Void sphere with $\sigma_\mathrm{f_0}\cos{\theta}$ surface charge imposed. See Appendix \ref{app:Field}, paragraph ``Graphs'' for interpreting Fig. \ref{fig:Situation 1ac}-\ref{fig:Situation 12}.}
	\label{fig:Situation 1ac}
\end{figure*}

[Note to editor and reviewers: this appendix, including figures \ref{fig:Situation 1ac} $...$ \ref{fig:Situation 2c COMSOL}, is written for educational purposes and to invite a critical review. It may be skipped in the final publication.]\\

\textbf{Pure dipole}\\
The field of a pure dipole $\boldsymbol{p}$, located at the origin of a spherical coordinate system and aligned with its vertical axis, is defined by voltage
\begin{equation}
	V=\frac{p}{4\pi\varepsilon_0}\cdot\frac{\cos{\theta}}{r^2} \label{eq:V_p}.
\end{equation}

\textbf{Spheroid dipole}\\
A dipole that is formed by distributing charge over the surface of a sphere with a $\sigma_0\cos{\theta}$ surface charge density produces that same voltage $V_2=V$ outside the sphere if
\begin{equation}
	p=\sigma_0\cdot \frac{4}{3}\pi R^3,
\end{equation}
so
\begin{equation}
	V_2=\frac{\sigma_0}{\varepsilon_0}\cdot\frac{1}{3}\cdot\frac{R^3\cos{\theta}}{r^2} 
	\label{eq:V_2_sigma_r_theta}.
\end{equation}
Thus, outside this spheroid dipole, the field is as if a pure dipole were placed at its center. This fulfills the boundary condition that far away the spheroid dipole should have a field like a pure dipole. Given the absence of other charge in the environment outside the sphere, this likeness must persist down to the sphere surface.\\

With $z=r\cos{\theta}$, in a compact mix of Cartesian and spherical coordinates:
\begin{equation}
	V_2=\frac{\sigma_0}{\varepsilon_0}\cdot\frac{1}{3}\cdot\frac{R^3}{r^3}\cdot z 
	\label{eq:V_2_sigma_r_z},
\end{equation}
Inside the sphere, a homogeneous, vertical field with voltage
\begin{equation}
	V_1=\frac{\sigma_0}{\varepsilon_0}\cdot\frac{1}{3}\cdot z
	\label{eq:V_1_sigma_r_z}
\end{equation} satisfies the condition that $V_1=V_2$ at the $r=R$ boundary.\\

It is important to consider the type of charge that the source terms $p$ and $\sigma_0\cos{\theta}$ represent. In Eq. (\ref{eq:V_p}) to (\ref{eq:V_1_sigma_r_z}), that is the net charge, which may consist of bound charge and free charge. Taking free charge $\sigma_\mathrm{f}=\sigma_\mathrm{f_0}\cos{\theta}$ as source term, the relations become
\begin{align}
	V_2&=\frac{\sigma_\mathrm{f_0}}{\varepsilon_0}\cdot\frac{1}{2\varepsilon_2+\varepsilon_1}\cdot\frac{R^3}{r^3}\cdot z \notag,\\
	V_1&=\frac{\sigma_\mathrm{f_0}}{\varepsilon_0}\cdot\frac{1}{2\varepsilon_2+\varepsilon_1}\cdot z \label{eq:V_12_sigma_f_r_z}.
\end{align}
where $\varepsilon_1$ and $\varepsilon_2$ denote the relative permittivity of the materials respectively inside and outside the sphere (and these materials are isotropic and respond linearly to the fields applied).\\

With $\boldsymbol{E}=-\nabla V$ the electric fields outside and inside the sphere become:
\begin{align}
	\boldsymbol{E}_2&=\frac{\sigma_\mathrm{f_0}}{\varepsilon_0}\cdot \frac{1}{2\varepsilon_2+\varepsilon_1}\cdot\boldsymbol{d}, \notag\\
	\boldsymbol{E}_1&=\frac{\sigma_\mathrm{f_0}}{\varepsilon_0}\cdot \frac{-1}{2\varepsilon_2+\varepsilon_1}\cdot\hat{\boldsymbol{z}}\label{eq:E12 given sigma_f_0}, 
\end{align}
with $\boldsymbol{d}$ as introduced in Eq. (\ref{eq:dipole_field_factor_d}).
\\

Note that homogeneous $\boldsymbol{E}_1$ and pure-dipole-like $\boldsymbol{E}_2$ both scale similarly if $\varepsilon_1$ or $\varepsilon_2$ is changed; their relative distribution stays the same.\\

\begin{figure*}[h!]
	\centering
	\begin{subfigure}[t]{0.49\textwidth}
		\centering
		\includegraphics[height=7.5cm]{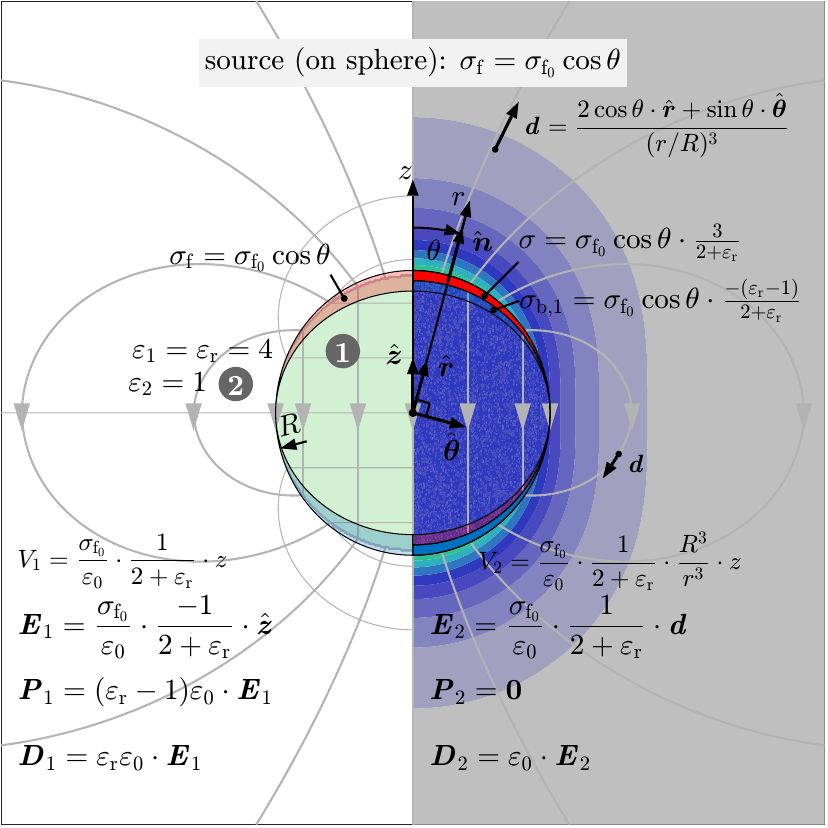}
		\caption{in vacuum}	
	\end{subfigure}%
	~
	\begin{subfigure}[t]{0.49\textwidth}
		\centering
		\includegraphics[height=7.5cm]{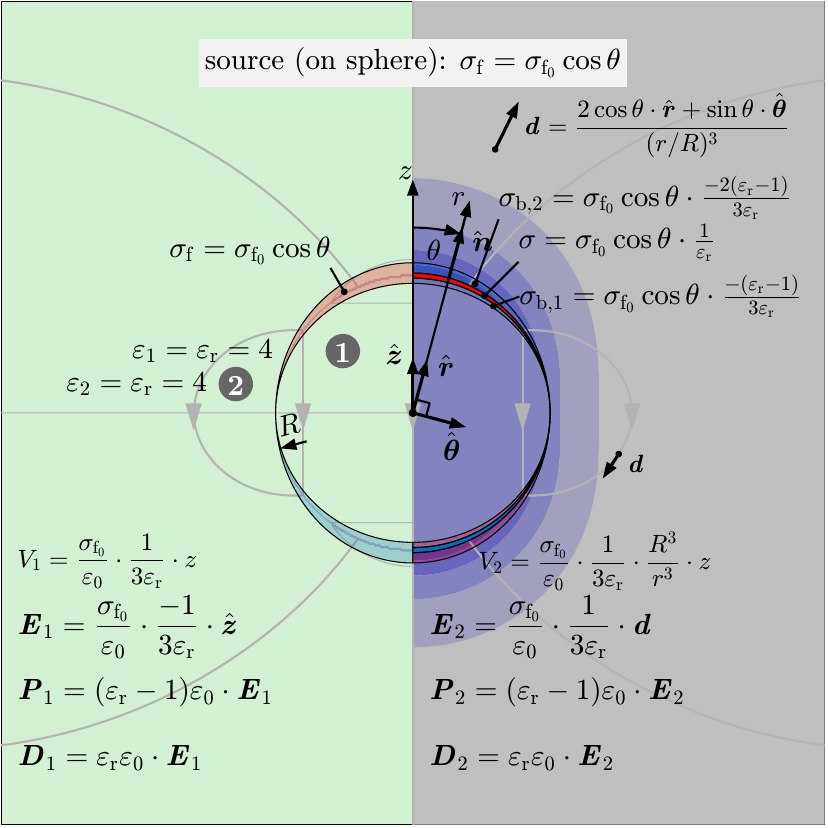}
		\caption{in dielectric}	
	\end{subfigure}
	\caption{Dielectric sphere with $\sigma_\mathrm{f_0}\cos{\theta}$ surface charge imposed.}
	\label{fig:Situation 1bd}
\end{figure*}

\begin{figure*}[h!]
	\centering
	\begin{subfigure}[t]{0.49\textwidth}
		\centering
		\includegraphics[height=7.5cm]{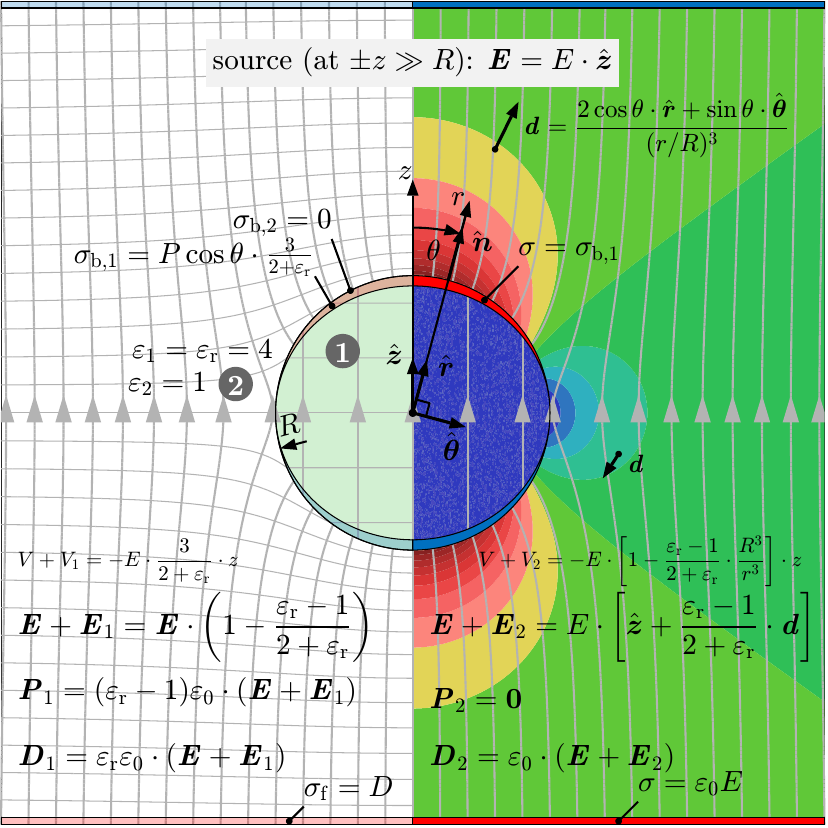}
		\caption{in vacuum}	
	\end{subfigure}%
	~
	\begin{subfigure}[t]{0.49\textwidth}
		\centering
		\includegraphics[height=7.5cm]{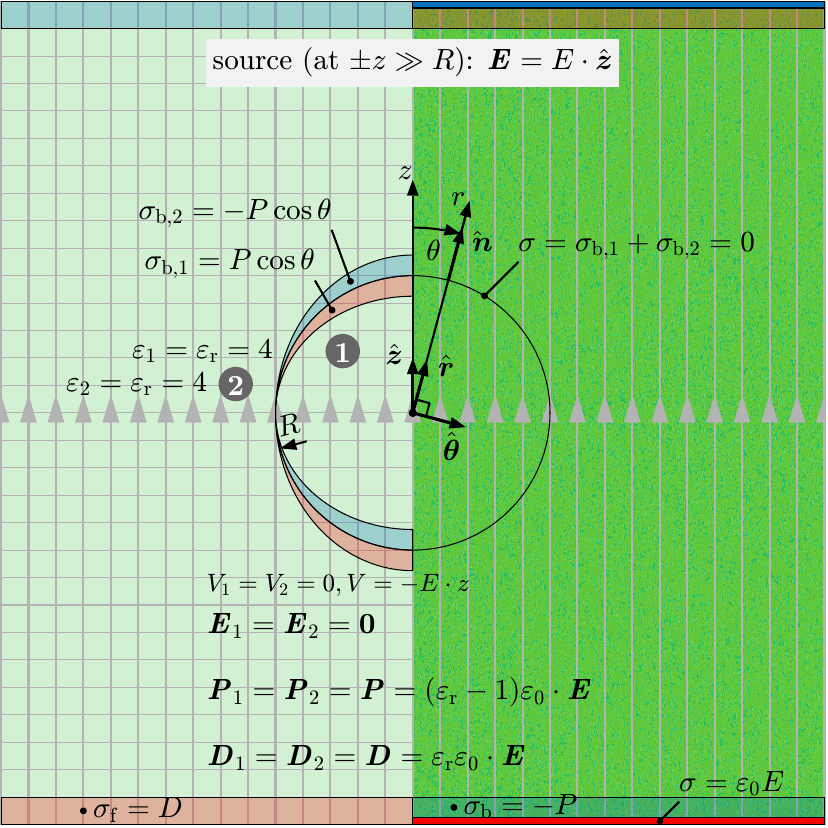}
		\caption{in same dielectric (so field and polarization are uniform)}
		\label{fig:Situation 2d}	
	\end{subfigure}
	\caption{Dielectric sphere with surrounding average field $\boldsymbol{E}$ imposed.}
	\label{fig:Situation 2bd}
\end{figure*}

\begin{figure*}[h!]
	\centering
	\begin{subfigure}[t]{0.49\textwidth}
		\centering
		\includegraphics[height=7.5cm]{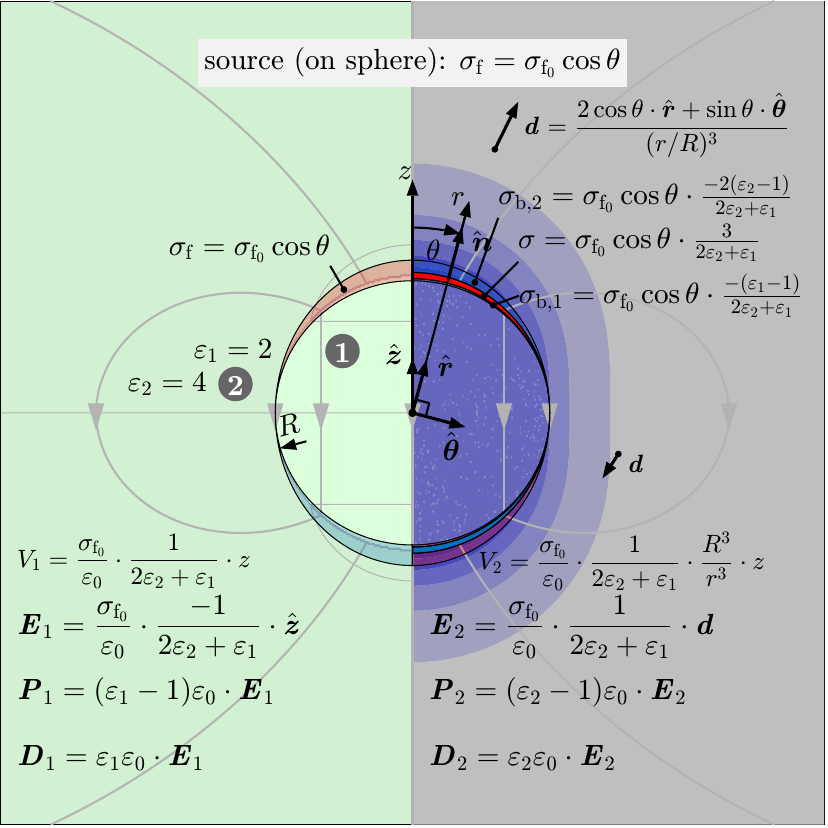}
		\caption{Any dielectric sphere in any dielectric with $\sigma_\mathrm{f_0}\cos{\theta}$ surface charge imposed.}	
		\label{fig:Situation 1}
	\end{subfigure}%
	~
	\begin{subfigure}[t]{0.49\textwidth}
		\centering
		\includegraphics[height=7.5cm]{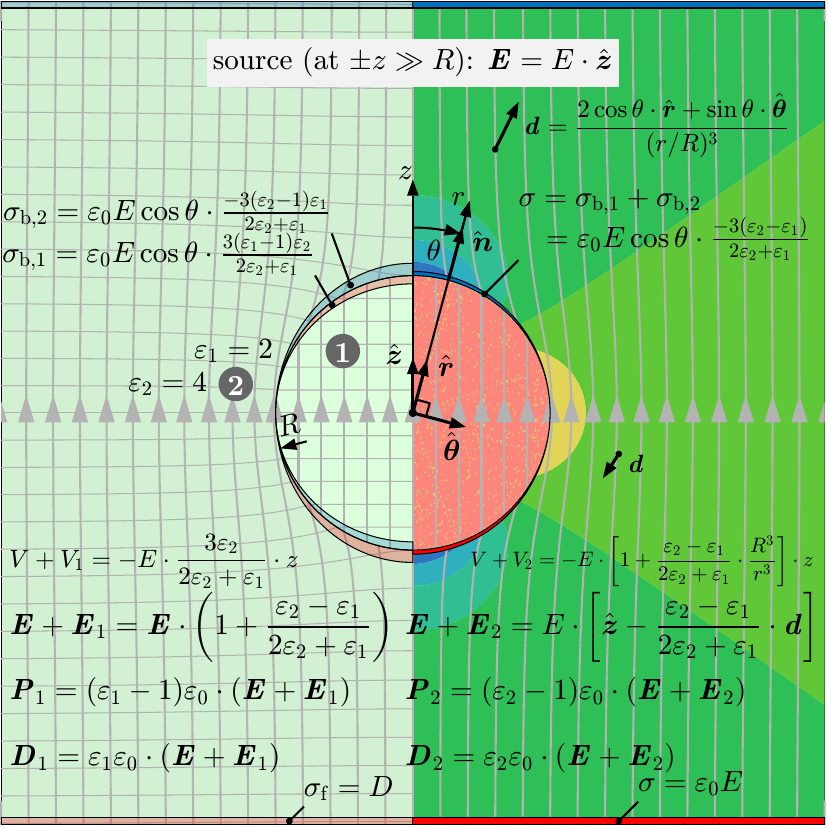}
		\caption{Any dielectric sphere in any dielectric with surrounding average field $\boldsymbol{E}$ imposed.}
		\label{fig:Situation 2}	
	\end{subfigure}
	\caption{Generic equations for any dielectric material inside/outside.}
	\label{fig:Situation 12}
\end{figure*}

\textbf{Graphs}\\
Situations with a void sphere are illustrated in Fig. \ref{fig:Situation 1ac} and with a dielectric sphere in Fig. \ref{fig:Situation 1bd}. Fig. \ref{fig:Situation 1} covers the generic solution. For these situations $\sigma_\mathrm{f}=P$, where $P$ is the polarization of the situation depicted in Fig. \ref{fig:Situation 2d} where $\varepsilon_2=\varepsilon_1=4$. The left-hand part of these and following graphs has white background color for $\varepsilon_r=1$ and light green for $\varepsilon_r>1$. The right-hand part's color scale, also used in Fig. \ref{fig:Situation 2c COMSOL}, quantifies field strength normalized to the average reference field strength $E$. Grey field lines contain arrows for their direction; their distance is a measure for inverse field strength\footnote{in all figures except Fig. \ref{fig:Situation 2c COMSOL}}. In the left half they are crossed by voltage contour lines. Surface charges are illustrated with an unrealistic non-zero thickness representing their local magnitude, opaque red/blue for positive/negative net charge $\sigma$ and transparent red/blue for $\sigma_\mathrm{f}$ and $\sigma_\mathrm{b}$. The surface charge responsible for $\boldsymbol{E}$ is shown at the top and bottom of the graphs, quite close to the spheres, but should be thought of so far away that the sphere field remains unaffected by its own mirror charge. The graphs also provide equations for $V$, $\boldsymbol{E}$, $\boldsymbol{P}$, $\boldsymbol{D}$, $\sigma$, $\sigma_\mathrm{f}$ and $\sigma_\mathrm{b}$ to allow comparing the uniform-field sphere solutions from all viewpoints.\\

\textbf{Sphere in otherwise average field}\\
Applying an average electric field $\boldsymbol{E}$ to a medium containing a sphere of different characteristics induces a $\cos{\theta}$-modulated surface charge onto its surface. One could say the sphere reacts by adding a dipole-like field to its environment and by changing the homogeneous field strength inside the sphere.\\

\begin{figure*}[h!]
	\centering
	\begin{subfigure}[t]{0.49\textwidth}
		\centering
		\includegraphics[height=7.5cm]{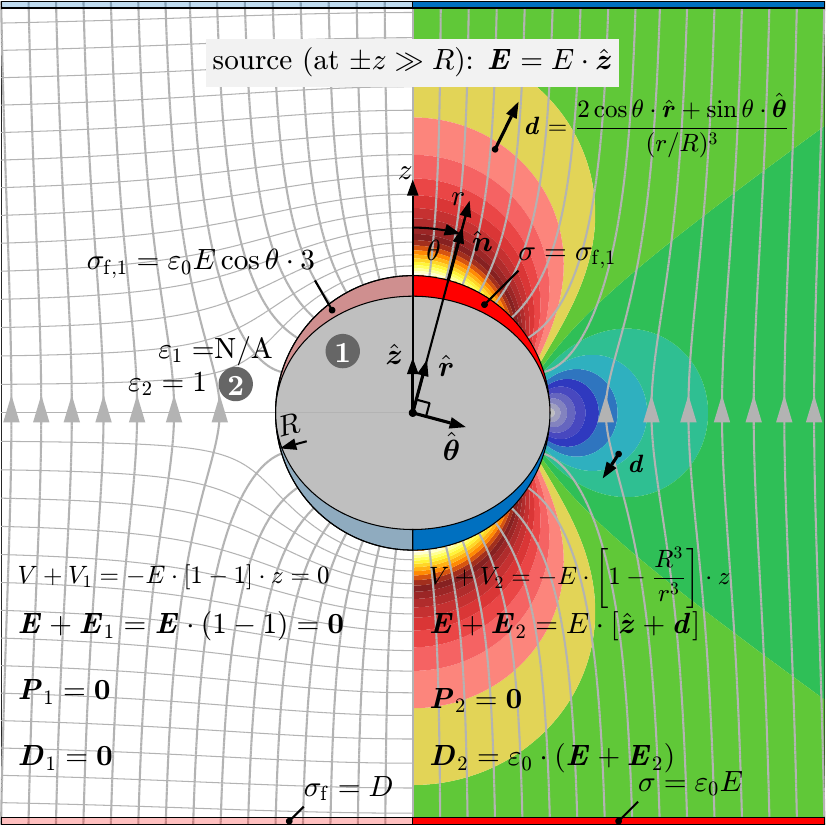}
		\caption{in vacuum}	
	\end{subfigure}%
	~
	\begin{subfigure}[t]{0.49\textwidth}
		\centering
		\includegraphics[height=7.5cm]{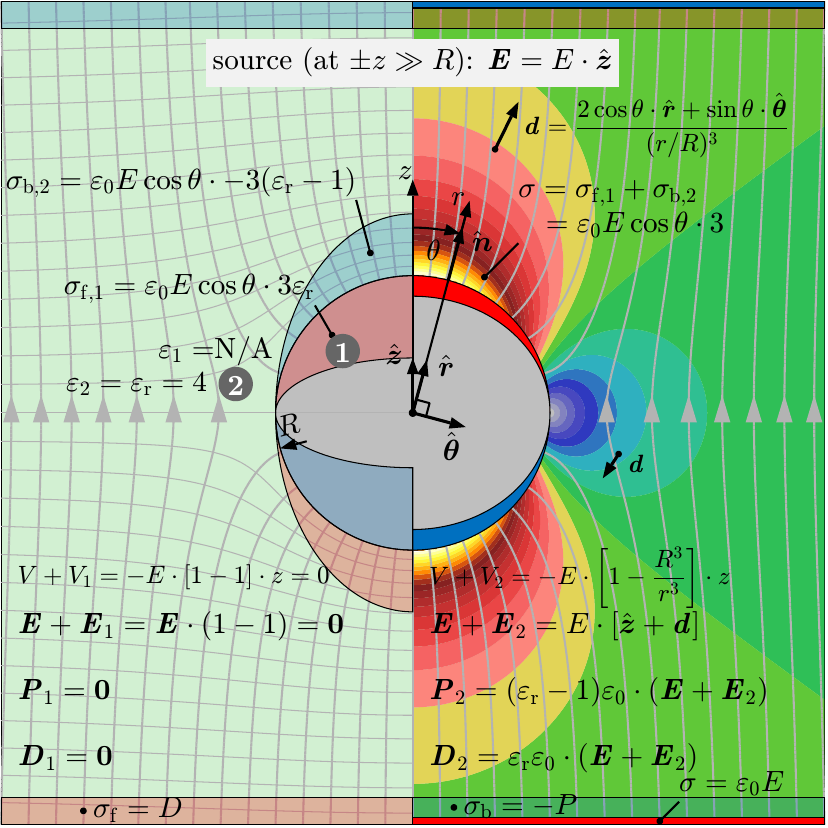}
		\caption{in dielectric}	
	\end{subfigure}
	\caption{Conductive sphere with surrounding average field $\boldsymbol{E}$ imposed.}
	\label{fig:Situation 2ef}
\end{figure*}

\textbf{- conductive sphere}\\
If the sphere is conductive (see Fig. \ref{fig:Situation 2ef}), free charge $\sigma_\mathrm{f}=3\varepsilon_2\cdot\varepsilon_0 E\cos{\theta}$ assembles inside at the sphere surface. If $\varepsilon_2>1$, the free charge induces bound charge $\sigma_\mathrm{b}=-3(\varepsilon_2-1)\varepsilon_2\cdot\varepsilon_0 E\cos{\theta}$ at the dielectric surface just around the sphere, which then counters part of the field from the free charge, leaving net charge $\sigma=3\cdot \varepsilon_0 E\cos{\theta}$. This generates field $\boldsymbol{E}_1$ which exactly cancels the imposed average field $\boldsymbol{E}$ leaving $\boldsymbol{E}+\boldsymbol{E}_1=\boldsymbol{0}$ inside the sphere. Outside, the net charge generates dipole field $\boldsymbol{E}_2$. The combined
\begin{equation}
	\boldsymbol{E}+\boldsymbol{E}_2=E\cdot\left(\hat{\boldsymbol{z}}+\boldsymbol{d}\right)
	\label{eq:Conductive sphere E+E2}
\end{equation}
peaks at $3E$ outside at the poles ($z=\pm R$) and is reduced to zero outside at the equator ($z=0$).\\

\begin{figure*}[h!]
	\centering
	\begin{subfigure}[t]{0.49\textwidth}
		\centering
		\includegraphics[height=7.5cm]{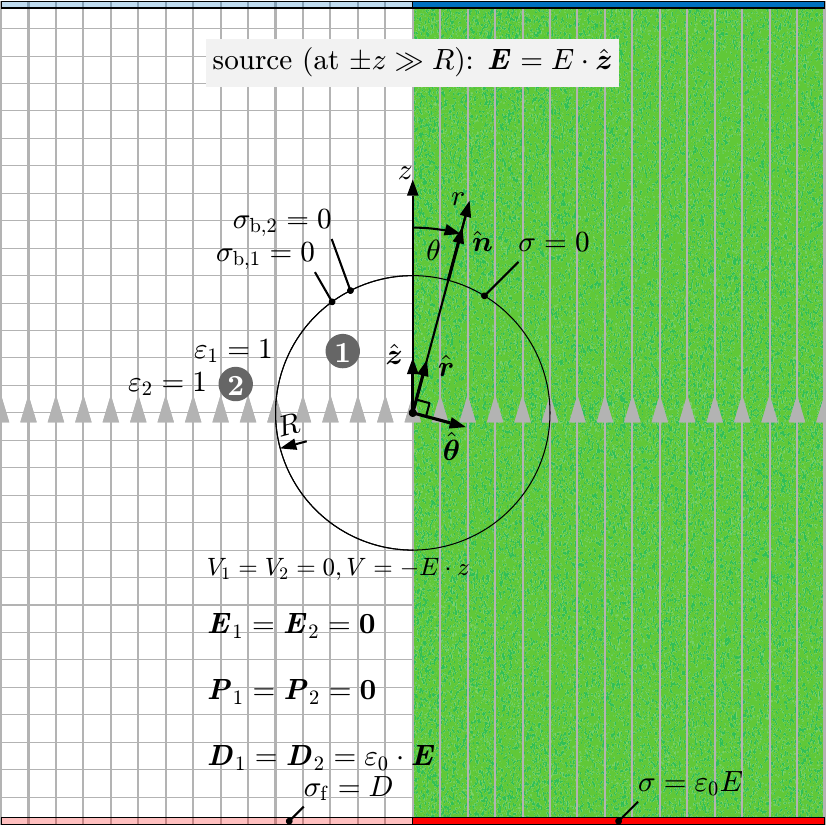}
		\caption{in vacuum}	
	\end{subfigure}%
	~
	\begin{subfigure}[t]{0.49\textwidth}
		\centering
		\includegraphics[height=7.5cm]{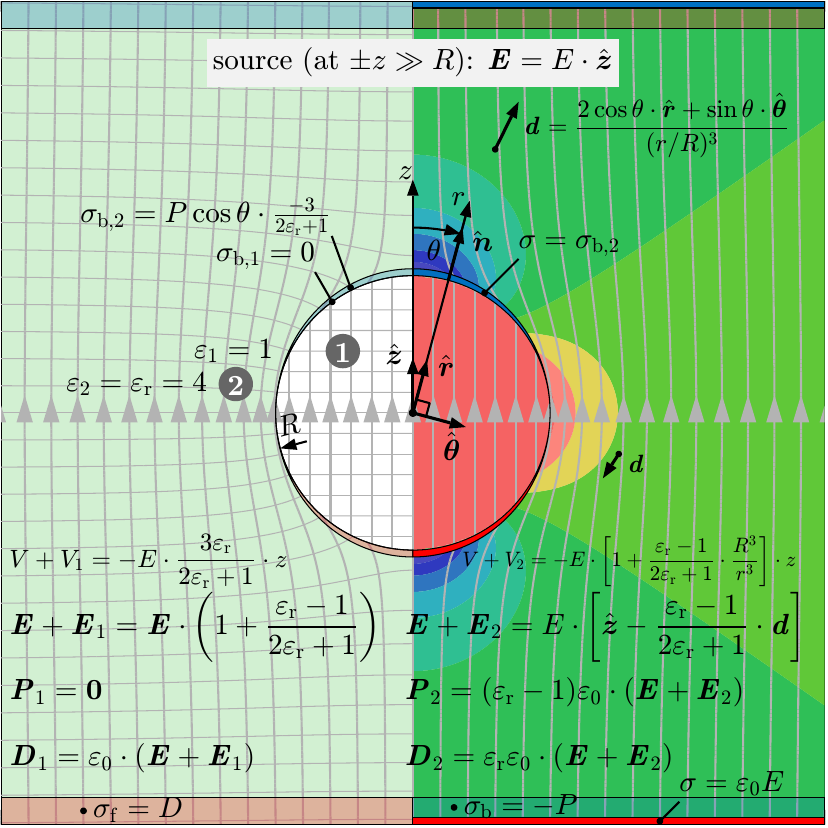}
		\caption{in dielectric}
		\label{fig:Situation 2c}	
	\end{subfigure}
	\caption{Void sphere with surrounding average field $\boldsymbol{E}$ imposed.}
	\label{fig:Situation 2ac}
\end{figure*}

\begin{figure}[h]
	\centering
	\includegraphics[width=\linewidth]{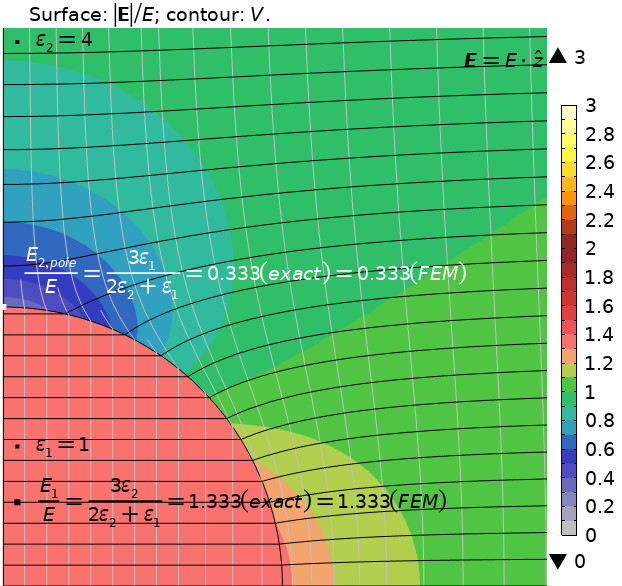}	
	\caption{Verification with finite element model, same situation as Fig. \ref{fig:Situation 2c}. The color scale is also used for figures \ref{fig:Situation 1ac} - \ref{fig:Situation 12}.}
	\label{fig:Situation 2c COMSOL}
\end{figure}

\textbf{- dielectric or void sphere}\\
If the sphere is void (Fig. \ref{fig:Situation 2ac}) or dielectric (Fig. \ref{fig:Situation 2bd}), a $\cos{\theta}$-modulated surface charge will still appear on the sphere surface, but it will now consist of bound charge in the dielectric boundary, with $\sigma=\sigma_\mathrm{b}=\frac{-3(\varepsilon_2-\varepsilon_1)}{2\varepsilon_2+\varepsilon_1}\cdot\varepsilon_0 E\cos{\theta}$ (see Eq. (\ref{eq:E+E_12 given E})). The field will still be a linear combination of the externally imposed average field plus the reaction field due to the (now bound) surface charge, adding the dipole field outside and enhancing or attenuating the vertical homogeneous field inside the sphere.\\

To find out how much surface charge the field $\boldsymbol{E}$ induces onto the sphere, Gauss's law for net charge on the sphere surface gives:
\begin{align}
	\varepsilon_0(E_{2,r}-E_{1,r})&=\sigma=\sigma_\mathrm{f}+\sigma_\mathrm{b}|_{r=R} \notag\\
	\sigma_\mathrm{f}=0\Rightarrow \varepsilon_0(E_{2,r}-E_{1,r})&=\sigma_\mathrm{b}|_{r=R} \label{eq:divE to sigma_b given E}
\end{align}
and Gauss's law for free charge on the sphere surface gives:
\begin{align}
	D_{2,r}-D_{1,r}=\sigma_\mathrm{f}|_{r=R},~~\sigma_\mathrm{f}&=0 \notag\\
	\Rightarrow \varepsilon_2\varepsilon_0(E_r+E_{2,r})-\varepsilon_1\varepsilon_0(E_r+E_{1,r})&=0|_{r=R}\label{eq:divD given E+E12}
\end{align}
and (\ref{eq:divE to sigma_b given E}) and (\ref{eq:divD given E+E12}) combine to solve:
\begin{align}
	\sigma_\mathrm{b}&=\varepsilon_0 E \cos{\theta}\cdot\frac{-3(\varepsilon_2-\varepsilon_1)}{2\varepsilon_2+\varepsilon_1}, \notag \\
	\boldsymbol{E}+\boldsymbol{E}_2&=E\cdot\left(\hat{\boldsymbol{z}}-\frac{\varepsilon_2-\varepsilon_1}{2\varepsilon_2+\varepsilon_1}\cdot\boldsymbol{d}\right), \notag \\
	\boldsymbol{E}+\boldsymbol{E}_1&=\boldsymbol{E}\cdot\left(1+\frac{\varepsilon_2-\varepsilon_1}{2\varepsilon_2+\varepsilon_1}\right), \label{eq:E+E_12 given E}
\end{align}
where $\boldsymbol{d}$ is the generic dipole field multiplier introduced in Eq. \ref{eq:dipole_field_factor_d}. The generic solution is shown in Fig. \ref{fig:Situation 2}. A finite element model with results in Fig. \ref{fig:Situation 2c COMSOL} confirms the analytical solution shown in Fig. \ref{fig:Situation 2c}.
The corresponding voltages are
\begin{align}
	V+V_2&=-E\cdot\left(1+\frac{\varepsilon_2-\varepsilon_1}{2\varepsilon_2+\varepsilon_1}\cdot\frac{R^3}{r^3}\right)\cdot z, \notag \\
	V+V_1&=-E\cdot\frac{3\varepsilon_2}{2\varepsilon_2+\varepsilon_1}\cdot z, \label{eq:V+V_12 given E}
\end{align}
Replacing, in Eq. (\ref{eq:V+V_12 given E}), coordinate $z$ by $r\cos{\theta}$, electric field strength $-E$ by magnetic field strength $I$, inverse dielectric constants $1/\varepsilon_1$ and $1/\varepsilon_2$ by "resistances" $k$ and $k'$ respectively, voltages $V+V_2$ and $V+V_1$ by potentials $P$ and $p_1$ respectively, and sphere radius $R$ by $a$:
\begin{align}
	P &= \left(Ir+A\frac{a^3}{r^2}\right) \cos{\theta}, A=\frac{k-k'}{2k+k'}I \nonumber \\ 
	p_1 &= Br \cos{\theta}, B=\frac{3k}{2k+k'}I \label{eq: Voltage cases 2 as per Maxwell}
\end{align}
which is Maxwell's exact formulation \cite{Maxwell-OFLF}.\\

\textbf{Derivation using Legendre/Laplacian}\\
An alternative method that leads to equal results is to solve Gauss's law $\varepsilon_0 \nabla \cdot\boldsymbol{E}=\sigma$. By definition $\boldsymbol{E}=-\nabla V$ so $\varepsilon_0 \nabla \cdot (-\nabla V)=\sigma$ which is also written as $\nabla^2 V=-\sigma/\varepsilon_0$ or $\Delta V=-\sigma/\varepsilon_0$, where $\Delta$ is called the Laplacian. In areas where $\sigma=0$ the field solution must satisfy $\Delta V=0$ and Legendre derived that for fields with axial symmetry polynomials in spherical coordinates with voltage
\begin{equation}
	V(r,\theta)=\sum_{\ell=0}^\infty \left(A_\ell r^\ell+\frac{B_\ell}{r^{\ell+1}}\right)P_\ell(\cos{\theta})\label{Legendre}
\end{equation}
satisfy $\Delta V=0$ for any set of fixed coefficients $A_\ell$ and $B_\ell$. Calculating a field solution then reverts to finding solutions which satisfy the voltage given at boundaries as well as the Legendre polynomial(s) evaluated at those boundaries, and then relying on the polynomials for the field between the boundaries, where there is no net charge.\\

For the field solutions in this paper it is enough to evaluate polynomials $P_0(\cos{\theta})=1$ and $P_1(\cos{\theta})=\cos{\theta}$; the higher terms describe fields for multipoles beyond the simple dipole needed here. Then,
\begin{itemize}
	\item $A_0=V_\mathrm{bias}$, can be set to any arbitrary value, $0$ in this paper,
	\item $B_0=\frac{q}{4\pi\varepsilon_0}$ models point charge $q$ placed at the origin, which can also be set to $0$ as there are no point charges involved in this paper,
	\item $A_1=-E$ models a homogeneous, vertically oriented field $\boldsymbol{E}=E\hat{\boldsymbol{z}}$, and
	\item $B_1=\frac{p}{4\pi\varepsilon_0}$ models a dipole, placed at the origin, with dipole moment oriented vertically $\boldsymbol{p}=p\hat{\boldsymbol{z}}$.
\end{itemize}

\textbf{Spheroid dipole}\\
For the case of impressing free surface charge $\sigma_\mathrm{f}=\sigma_\mathrm{f_0}\cos{\theta}$ onto a sphere of relative permittivity $\varepsilon_1$ in medium with $\varepsilon_2$, coefficients $A_1$ and $B_1$ are used outside the sphere and, similarly $C_1$ and $D_1$ inside the sphere. $A_1=0$ since the spheroid dipole will not create an average field. $D_1=0$ since there is no dipole to generate a dipole field inside. To find $B_1$ and $C_1$ the first boundary condition to satisfy is Gauss's law in matter on the sphere surface:
\begin{align}
	\sigma_\mathrm{f}&=(D_{2,r}-D_{1,r})|_{r=R} \notag\\
	\Rightarrow \sigma_\mathrm{f}&=(\varepsilon_2\varepsilon_0 E_{2,r}-\varepsilon_1\varepsilon_0 E_{1,r})|_{r=R} \notag\\
	\Leftrightarrow \sigma_\mathrm{f}&=\left.\left(\varepsilon_2\varepsilon_0\cdot -\frac{\partial}{\partial r}\left(B_1\frac{\cos{\theta}}{r^2}\right)\right)\right|_{r=R} \notag\\
	&-\left.\left(\varepsilon_1\varepsilon_0\cdot -\frac{\partial}{\partial r}\left(C_1 r\cos{\theta}\right)\right)\right|_{r=R} \notag \\
	\Rightarrow \sigma_\mathrm{f}&=\sigma_\mathrm{f_0}\cos{\theta}=\varepsilon_2\varepsilon_0\cdot 2B_1\frac{\cos{\theta}}{R^3} \notag \\
		&~~~~~~~~~~~~~~~+\varepsilon_1\varepsilon_0\cdot C_1\cos{\theta} \notag \\
	\Leftrightarrow \frac{\sigma_\mathrm{f_0}}{\varepsilon_0}&=2\varepsilon_2\cdot\frac{B_1}{R^3}+\varepsilon_1\cdot C_1 
	\label{eq:divD given E12 spheroid dipole}.
\end{align}

The second boundary condition is that the voltages inside and outside must match at $r=R$:
\begin{align}
	B_1\frac{\cos{\theta}}{R^2}&=C_1 R cos{\theta} \notag \\
	\Leftrightarrow B_1&=C_1 R^3
	\label{eq:V1=V2 at R spheroid dipole}.
\end{align}
Eq. (\ref{eq:divD given E12 spheroid dipole}) and (\ref{eq:V1=V2 at R spheroid dipole}) combine to
\begin{align}
	B_1&=\frac{\sigma_\mathrm{f_0}}{\varepsilon_0}\cdot\frac{R^3}{2\varepsilon_2+\varepsilon_1}, \notag \\
	C_1&=\frac{\sigma_\mathrm{f_0}}{\varepsilon_0}\cdot\frac{1}{2\varepsilon_2+\varepsilon_1}
	\label{eq:B1 and C1 spheroid dipole given sigma_f}.
\end{align}
These coefficients confirm the field of Eq. (\ref{eq:E12 given sigma_f_0}) derived before.\\

\textbf{Sphere in otherwise average field}\\
For the case of impressing average field $\boldsymbol{E}$ in the medium containing a sphere, and using the same coefficients, a similar approach is used. Now, $A_1=-E$ since the average outside field has to match the imposed average field. $D_1=0$ since also here is no dipole to generate a dipole field inside. Again, the first boundary condition to solve is Gauss's law in matter at the sphere surface:
\begin{align}
	\sigma_\mathrm{f}=(D_{2,r}-D_{1,r})|_{r=R}&=0 \notag\\	
	\Rightarrow \left.\left(\varepsilon_2\varepsilon_0 (E_r+E_{2,r})\right)\right|_{r=R}-& \notag \\
		     \left.\left(\varepsilon_1\varepsilon_0 (E_r+E_{1,r})\right)\right|_{r=R}&=0 \notag \\
	\Leftrightarrow \varepsilon_2\varepsilon_0\cdot\left(E\cos{\theta}-\left.\frac{\partial}{\partial r}\left(B_1\frac{\cos{\theta}}{r^2}\right)\right|_{r=R}\right)-& \notag \\
			  \varepsilon_1\varepsilon_0\cdot\left(E\cos{\theta}-\left.\frac{\partial}{\partial r}\left(C_1 r\cos{\theta}\right)\right|_{r=R}\right)&=0 \notag \\
	\Leftrightarrow (\varepsilon_2-\varepsilon_1)\cdot E\cos{\theta}+\varepsilon_2\cdot 2B_1\frac{\cos{\theta}}{R^3}+& \notag \\
	\varepsilon_1\cdot C_1\cos{\theta}&=0 \notag \\
	\Leftrightarrow 2\varepsilon_2\cdot \frac{B_1}{R^3}+\varepsilon_1\cdot C_1=-(\varepsilon_2-\varepsilon_1)\cdot E&. \notag \\ 
	\label{eq:divD given E+E12 Legendre}
\end{align}
Again, the second boundary condition is that the voltages inside and outside must match at $r=R$:
\begin{align}
	B_1\frac{\cos{\theta}}{R^2}-Ez&=C_1 R cos{\theta}-Ez \notag \\
	B_1&=C_1 R^3
	\label{eq:V1=V2 at R with E imposed}.
\end{align}
Eq. (\ref{eq:divD given E+E12 Legendre}) and (\ref{eq:V1=V2 at R with E imposed}) combine to
\begin{align}
	B_1&=-E\cdot R^3\frac{\varepsilon_2-\varepsilon_1}{2\varepsilon_2+\varepsilon_1}, \notag \\
	C_1&=-E\cdot\frac{\varepsilon_2-\varepsilon_1}{2\varepsilon_2+\varepsilon_1}
	\label{eq:B1 and C1 spheroid dipole given E}.
\end{align}
These coefficients confirm the field of Eq. (\ref{eq:E+E_12 given E}) and (\ref{eq:V+V_12 given E}) derived before.\\

\bibliographystyle{plain} 
\bibliography{UFS} 

\begin{thebibliography}{10}

\bibitem{Arndt-Hummel-GRFAtDSG}
J.~Arndt and W.~Hummel.
\newblock {The general refractivity formula applied to densified silicate
  glasses}.
\newblock {\em Physics and Chemistry of Minerals 15}, pages 363--369, 1988.

\bibitem{Ashcroft}
Neil Ashcroft and David Mermin.
\newblock {\em {Solid State Physics}}.
\newblock Saunders College Publishing, 1st edition, 1976.
\newblock p.541.

\bibitem{Clausius-MBdE}
Rudolf Clausius.
\newblock {\em {Die mechanische Behandlung der Electricität}}.
\newblock Springer, 2nd edition, 1879.
\newblock p.94.

\bibitem{Eremin}
I.E. Eremin et~al.
\newblock {System modification of the equation
  Lorenz-Lorentz-Clausius-Mossotti}.
\newblock {\em International Journal for Light and Electron Optics}, 2021.

\bibitem{FLPII-11}
Richard Feynman.
\newblock {\em {The Feynman lectures on physics, Chapter 11. Inside
  Dielectrics}}, volume~II.
\newblock Addison-Wesley, 13 edition, 1964.

\bibitem{Griffiths-ItE}
David~J. Griffiths.
\newblock {\em {Introduction to electrodynamics}}.
\newblock Pearson, 4th edition, 2013.
\newblock Ex. 4.2, problem 4.41.

\bibitem{Jackson-CE}
John~David Jackson.
\newblock {\em {Classical Electrodynamics}}.
\newblock Pearson, 1st edition, 1962.
\newblock p.115-116.

\bibitem{Kittel}
Charles Kittel.
\newblock {\em {Introduction to Solid State Physics}}.
\newblock John Wiley \& Sons, 8th edition, 2005.
\newblock p.460.

\bibitem{Lobanov-ESMPSG}
Sergey Lobanov, Sergio Speziale, et~al.
\newblock {Electronic, structural, and mechanical properties of SiO$_2$ glass
  at high pressure inferred from its refractive index}.
\newblock {\em Physical Review Letters}, 128(077403), 2022.

\bibitem{Lorentz-UdBzdFdLudK}
Hendrik Lorentz.
\newblock {Ueber die Beziehung zwischen der Fortpflanzungsgeschwindigkeit des
  Lichtes und der Körperdichte}.
\newblock {\em Annalen der Physik}, pages 641--665, 1880.

\bibitem{Lorenz-EoTUoLB}
Ludvig Lorenz.
\newblock {Experimentale og theoretiske undersøgelser over legemers
  brydningsforhold}.
\newblock {\em Det kongelige danske Videnskabernes Selskabs Skrifter 5(8)},
  page 203–248, 1869.

\bibitem{Maxwell-OFLF}
James~Clerk Maxwell.
\newblock {On Faraday's lines of force}.
\newblock {\em Transactions of the Cambridge Philosophical Society}, x(Part
  1):70--71, 1858.

\bibitem{Mossotti-MDhSDdE}
Ottaviano-Fabrizio Mossotti.
\newblock {Discussione analytica sull'influenza che l'azione di un mezzo
  dielettrico ha sulla distribuzione dell'electricità alla superficie di più
  corpi elettrici disseminati in esso}.
\newblock {\em Memorie di matematica e di fisica delle Società italiana delle
  scienze, tomo XXIV, parte seconda}, pages 1--26, 1846.

\bibitem{Onsager-EMML}
Lars Onsager.
\newblock {Electric Moments of Molecules in Liquids}.
\newblock {\em Journal of the American Chemical Society}, pages 1486--1493,
  1936.

\bibitem{Schwerdtfeger-Nagle}
Peter Schwerdtfeger and Jeffey~K. Nagle.
\newblock {2018 Table of static dipole polarizabilities of the neutral elements
  in the periodic table}.
\newblock {\em Molecular Physics}, 2018.

\bibitem{TanArndtXie-OPoDSG}
C.Z. Tan, J.~Arndt, and H.S. Xie.
\newblock {Optical properties of densified silica glasses}.
\newblock {\em Physica B}, (252):28--33, 1998.

\bibitem{Zangwill}
Andrew Zangwill.
\newblock {\em {Modern Electrodynamics}}.
\newblock Cambridge University Press, 1st edition, 2012.
\newblock p.174-175.

\end{thebibliography}

\end{document}